\documentclass[10pt,journal,  oneside, twocolumn]{IEEEtran}

\usepackage{multirow}
\usepackage{latexsym}
\usepackage{graphicx}
\usepackage{float}
\usepackage{amsmath}
\usepackage{amsthm}
\usepackage{lipsum}
\usepackage{subfig}
\usepackage{graphicx}
\usepackage{authblk}
\usepackage{bm}
\usepackage{booktabs}
\usepackage{amsthm}
\usepackage[section]{placeins}
\usepackage{soul}

\usepackage{colortbl}

\usepackage{enumitem}

\usepackage{CJKutf8}

\makeatletter  
\newif\if@restonecol  
\makeatother

\usepackage[linesnumbered,ruled,vlined]{algorithm2e}
\usepackage{algpseudocode}  
\usepackage{amsmath}

\usepackage{amssymb}

\usepackage{mathrsfs}
\usepackage{subfig}
\usepackage{caption}
\newcounter{problem}

\SetKwInput{KwInitialize}{Initialize}

\renewcommand{\baselinestretch}{1.0}

\begin{document}

\title{Network Slice-based Low-Altitude Intelligent Network for Advanced Air Mobility}


\author{Kai Xiong~\IEEEmembership{Member,~IEEE}, Yutong Chen, Supeng Leng~\IEEEmembership{Member,~IEEE}, Chau Yuen,~\IEEEmembership{Fellow,~IEEE} 


\thanks{

K. Xiong and Y. Chen and S. Leng are with School of Information and Communication Engineering, University of Electronic Science and Technology of China, Chengdu, 611731, China; and, Shenzhen Institute for Advanced Study, University of Electronic Science and Technology of China, Shenzhen, 518110, China.
}



\thanks{
 C. Yuen is with School of Electrical and Electronics Engineering, Nanyang Technological University, 639798, Singapore. 
 }

\thanks{The financial support of National Natural Science Foundation of China (NSFC), Grant No.62201122.
}


}


\maketitle

\begin{abstract}
Advanced Air Mobility (AAM) is transforming transportation systems by extending them into near-ground airspace, offering innovative solutions to mobility challenges. In this space, electric vertical take-off and landing vehicles (eVTOLs) perform a variety of tasks to improve aviation safety and efficiency, such as collaborative computing and perception. However, eVTOLs face constraints such as compacted shape and restricted onboard computing resources. These limitations necessitate task offloading to nearby high-performance base stations (BSs) for timely processing.
Unfortunately, the high mobility of eVTOLs, coupled with their restricted flight airlines and heterogeneous resource management creates significant challenges in dynamic task offloading.
To address these issues, this paper introduces a novel network slice-based Low-Altitude Intelligent Network (LAIN) framework for eVTOL tasks.
By leveraging advanced network slicing technologies from 5G/6G, the proposed framework dynamically adjusts communication bandwidth, beam alignment, and computing resources to meet fluctuating task demands. Specifically, the framework includes an access pairing method to pre-schedule optimal eVTOL-BS-slice assignments, a pre-assessment algorithm to avoid resource waste, and a deep reinforcement learning-based slice orchestration mechanism to optimize resource allocation and lifecycle management.
Simulation results demonstrate that the proposed framework outperforms existing benchmarks in terms of resource allocation efficiency and operational/violation costs across varying eVTOL velocities. This work provides valuable insights into intelligent network slicing for future AAM transportation systems.

\end{abstract}


\IEEEpeerreviewmaketitle

\section{Introduction}

\IEEEPARstart{A}{s} urban traffic congestion continues to worsen, the need for innovative transportation solutions has become more urgent than ever. Advanced Air Mobility (AAM), powered by electric vertical take-off and landing (eVTOL) technology, offers a promising approach to alleviate ground congestion and improve urban commuting \cite{10388419Henglai}.

According to the National Aeronautics and Space Administration (NASA), eVTOLs must support a variety of applications in low-altitude airspace. These include real-time obstacle avoidance, which requires precise flight planning based on environmental data; emergency handling, which involves the rapid transmission of danger alerts to control centers; and air traffic management, which coordinates flight airlines for multiple eVTOLs through air-ground communication systems \cite{{Ehang3248},{10522499Feng}}. All of these applications rely on efficient computing systems with substantial computational resources \cite{9405488Lyu}. Due to these applications, eVTOLs will generate large amounts of data, requiring low-altitude networks to share individual processing results for intelligent collaborations.

To address these needs, the Low-Altitude Intelligent Network (LAIN) has been developed as a communication system designed to support diverse low-altitude tasks, such as low latency, high throughput, and strong reliability. 
LAIN aims to establish highly customized connections between manned eVTOLs, unmanned aerial vehicles (UAVs), and ground infrastructure, using advanced communication, computing, and flight control technologies \cite{{9814972Yang},{Pan10278101},{9714482Ziye}}.

However, designing the LAIN system is particularly challenging due to the stringent size, battery, and mobility constraints of eVTOLs. These challenges must be addressed alongside the diverse services from the AAM ecosystem. 
Additionally, the LAIN framework also must consider the unique trajectory of eVTOLs, such as vertically layered airlines that benefit air traffic safety and capacity of the low-altitude airspace \cite{42358726Sunil}. 
While the vertical lift and horizontal cruising along the layered airlines significantly affect the LAIN performance and design.


To overcome these challenges, we leverage network slicing (NS) technology from 5G/6G networks. It enables the dynamic segmentation and allocation of heterogeneous resources based on multiple task-specific requirements within the LAIN framework. By integrating multiple infrastructures into a virtual resource pool, NS can orchestrate overall heterogeneous resources to maximize system efficiency \cite{{9144143Caballero},{10082939Seid}}. 
Moreover, combined with the characteristics of the low-altitude airspace, the low-altitude NS has to be capable of dynamically adjusting bandwidth, air beam alignment, and computing resources to accommodate airline constraints and task demands of eVTOLs.


Consequently, this paper introduces a novel NS-based LAIN framework aimed at improving AAM operational efficiency. The framework optimizes dynamic task offloading by considering factors such as the task lifecycle, BS/eVTOL position/velocity, specific layered airline constraints, and heterogeneous resource allocations. Additionally, we propose a slice admission control module that pre-establishes the eVTOL-BS-slice pairing and pre-assigns available resources for task offloading. This module enhances resource utilization and reduces overall system costs. The main contributions are summarized as follows:

\begin{itemize}

\item We propose a low-altitude network slice tailored to highly dynamic eVTOL tasks. This slice undergoes a complete lifecycle evolution of "initialization—scaling—disposal," driven by flight dynamics and task requirements. Specifically, we developed an intelligent slicing orchestration that leverages multi-agent deep reinforcement learning to account for factors such as specific flight airlines, antenna beam alignment, and slice lifecycle management. 



\item We develop an eVTOL flight model for the layered AAM airspace, which involves ascending through two flight layers: a low-speed layer and a high-speed layer. eVTOLs in different layers exhibit distinct flight behaviors, as well as varying task and resource requirements. The proposed low-altitude network slice can dynamically adjust heterogeneous resources to accommodate the different needs of eVTOLs in these layers.


\item We design a novel slice admission control module that includes both access pairing and resource pre-assessment algorithms. The access pairing algorithm selects the optimal eVTOL-BS-slice pair, while the pre-assessment algorithm prevents resource waste by pre-assigning the total amount of resources. These algorithms account for eVTOL mobility and preemptively reject task offloading likely to fail, based on eVTOL dynamics and task requirements. By performing slice admission control prior to slice orchestration, this approach reduces the complexity of resource management, minimizing operational costs and enhancing system efficiency.



\end{itemize}

The remainder of this paper is organized as follows. Section II reviews the related works. Section III presents the whole system architecture. Section IV provides the simulation results and the performance discussion. Finally, we conclude in Section V.

\section{Related Work}
The development of eVTOL has gained significant attention from engineers and researchers as a key component for Advanced Air Mobility (AAM) transportation systems. 
Here, safe eVTOL aviation heavily relies on an efficient low-altitude network.
Based on this low-altitude network, eVTOL will generate diverse aerial applications, which further increases the requirements for fine scheduling of computing, communication, and flight control.
Since network slicing is a critical enabler of 5G/6G networks that can simultaneously guarantee the Quality of Service (QoS) of multiple applications, it naturally associates network slicing technology with low-altitude networks to improve overall AAM operational efficiency.
In the following subsections, we review the research progress in both AAM and network slicing.


\subsection{Advanced Air Mobility}

As an attractive game-changer, AAM offers flexible and efficient mobility options in urban environments \cite{{9447255Adam},{Choi10393649}}. 
Aircraft such as eVTOLs and UAVs are expected to serve various roles, including air taxis, metro alternatives, and last-mile delivery services \cite{9952882Kim}. According to a report by Morgan Stanley, the global AAM market is projected to reach \$1.5 trillion by 2040, rivaling the potential market of autonomous vehicles \cite{Morgan2019}.

Current research on AAM focuses on two primary aspects: eVTOL mobility management and low-altitude network design \cite{Choi10393649}. For example, Zhou \textit{et al.} \cite{9983486Zhou} developed a dual-layer optimization framework that integrates path planning, transmission power control, and air-to-ground transmission scheduling. This framework aims to minimize UAV flight energy consumption while ensuring high-reliability communication. Similarly, Sinha \textit{et al.} \cite{10620635Sinha} proposed a communication scheme specifically designed for high-mobility environments. Piccioni \textit{et al.} \cite{10504640Piccioni} introduced an enhanced air-to-ground access algorithm that ensures high connectivity, low-cost, and instant network access for UAVs while reducing dependence on ground BSs.

However, these studies primarily focus on eVTOL/UAV-based network optimization without fully considering the unique airspace structure of AAM systems. Sunil \textit{et al.} \cite{42358726Sunil} confirmed that the vertical layered airspace improves both air traffic safety and capacity. eVTOLs, operating within AAM, must adhere to a specific layered airspace structure, which has significant implications for low-altitude network design. Furthermore, these studies often fail to address the coordinated management of heterogeneous resources, such as communication and computation, within low-altitude environments.

However, the above studies on eVTOL/UAV-based network optimization overlook the unique airspace structure of AAM systems. 
Sunil \textit{et al.} \cite{42358726Sunil} confirmed that the vertical layered airspace benefits both air traffic safety and capacity. eVTOLs, operating within AAM, must adhere to a specific layered airspace structure, which has significant implications for low-altitude network design. Furthermore, these studies often fail to address the coordinated management of heterogeneous resources, such as communication and computation, within low-altitude environments.






\subsection{Network Slicing}
In the context of UAV-assisted wireless networks, various network slice solutions have been proposed. 
Peng \textit{et al.} \cite{9453787Peng} explored a proactive network slicing strategy based on location prediction, which mitigates the mismatch between slice supply and demand by proactively managing the UAV network. However, the slice creation and configuration processes in this approach are time-consuming. To avoid frequent reconfigurations, Wei \textit{et al.} \cite{10679214Wei} proposed an intelligent hierarchical UAV network slicing framework that operates on different time scales. 
This framework allows for adaptive control, reducing unnecessary reconfigurations and improving operational efficiency while minimizing resource consumption.

Cho \textit{et al.} \cite{9289473Cho} developed a UAV RAN slicing resource allocation algorithm designed to accommodate both enhanced mobile broadband (eMBB) and massive machine-type communications (mMTC) users. While these studies offer efficient and fair strategies for resource allocation, they do not fully consider the mobility of UAVs. Bellone \textit{et al.} \cite{10257231Bellone} introduced a reinforcement learning-based bandwidth allocation strategy that accounts for UAV mobility. However, this approach is primarily designed for a single resource optimization in terrestrial 5G networks, where UAVs are limited to movement on a fixed-altitude 2D plane, significantly limiting their mobility.
Therefore, a significant gap remains in the development of network slicing solutions specifically tailored for low-altitude airspace.

To address this gap, this paper proposes an intelligent network slicing solution for low-altitude environments. This method takes into account the unique flight patterns of eVTOLs operating in 3-Dimensional low-altitude airspace. 
It introduces a slice admission control module that matches the optimal ground access BS based on the mobility status of eVTOLs. Additionally, a task pre-assessment method is proposed to reduce overall resource overhead. 
To handle the dynamic nature of tasks caused by eVTOL mobility, a slice lifecycle management mechanism is introduced. This mechanism, which includes initialization, reconfiguration, and disposal phases, adapts to the high dynamics of eVTOL tasks, thereby maximizing system resource utilization efficiency.

\section{System Architecture}
\begin{figure}[h]
\centering
     \includegraphics[width=.48\textwidth]{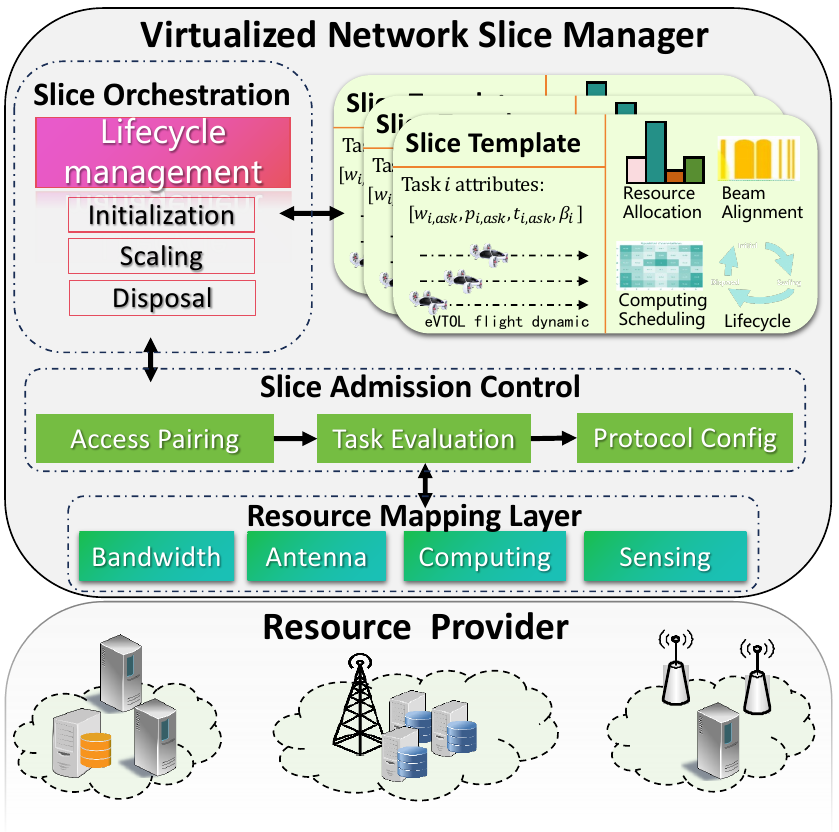} 
\caption{Intelligent Network Slice Framework.} 
\label{frameworkness}
\end{figure}

This paper proposes a novel network slice-based AAM system designed to provide efficient network services for various eVTOL applications in AAM. 
As shown in Fig.~\ref{frameworkness}, the low-altitude network slice is managed by the Virtualized Network Slice Manager (VNSM). 
This proposed VNSM consists of three key modules: the resource mapping layer, slice admission control, and slice orchestration.

Wherein the resource mapping layer abstracts heterogeneous resources, including virtual communication, computing, and sensing resources. 
These virtualized resources simplify the invocation of heterogeneous resources, such as bandwidth, antenna beams, and high-performance chips, making it easier to efficiently support multiple task payloads for eVTOLs.

Furthermore, the proposed slice admission control module generates the eVTOL access strategy, evaluates task complexity, and configures the network protocol. 
Its operation follows a sequential process: first, it identifies the optimal eVTOL-BS-Slice pairs based on dynamic information (e.g., eVTOL position, velocity) to maximize task offloading efficiency. 
Next, this module pre-schedules an appropriate total volume of resources to avoid resource waste. 
Finally, after determining the eVTOL-BS-Slice pairs, the slice admission control module selects suitable network protocols based on the task requirements.

The slice orchestration module operates on the top of the resource mapping layer and slice admission control module.
It is responsible for the lifecycle management of slice templates, which includes resource allocation during slice initialization, scaling, and disposal.

In detail, this module generates multiple slice templates and corresponding resource allocation schemes for each lifecycle stage. The initialization stage corresponds to the start of task offloading, the scaling stage handles dynamic task changes, and the disposal stage marks task completion. Using templates enables pre-determined resource allocation, protocol configurations, and service flows for network slices at various stages, allowing the system to quickly respond to dynamic eVTOL task demands.

\begin{figure}[h]
\centering
     \includegraphics[width=.48\textwidth]{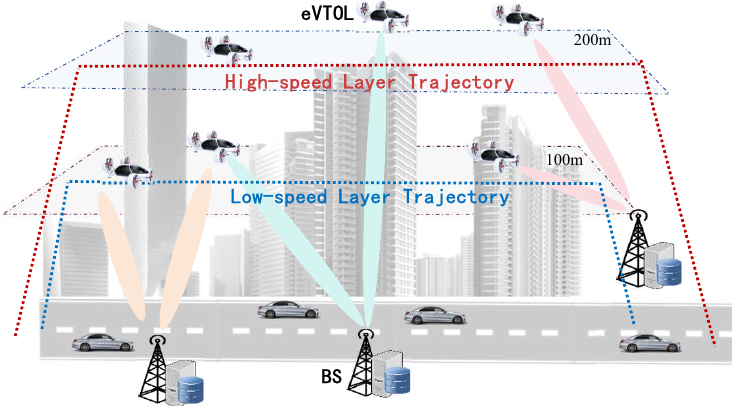} 
\caption{Aviation behaviors of eVTOLs in AAM.} 
\label{AAM_Scenario}
\end{figure}

Next, we investigate the layered AAM scenario. Sunil \textit{et al.} \cite{42358726Sunil} demonstrated that vertically layered airspace enhances air traffic safety and capacity. 
Accordingly, Fig.~\ref{AAM_Scenario} depicts the structure of a layered UAM system, where adjacent vertical layers are separated by a height of $100$ m. To maintain traffic safety within the layered airspace, all eVTOLs operating within the same layer must fly at the same velocity, and the prescribed velocity of the layer raises with the layer altitude.
Therefore, short-range eVTOLs can operate in lower altitude layers, also known as the low-speed layers. In contrast, long-range eVTOLs, which require higher speeds, can fly at higher altitudes in the high-speed layers. These layered flight settings minimize relative velocity between eVTOLs within the same layer and improving overall traffic efficiency.

Within the layered AAM, multiple eVTOLs and BSs collaborate to complete low-altitude tasks. For example, several eVTOLs can work together to enhance task execution efficiency and accuracy through cooperative sensing. However, after merging the sensing data into the lead eVTOL, the onboard computing and storage resources of individual eVTOLs may be insufficient to process the sensing fusion locally. In such cases, the eVTOL can offload the processing task to a ground BS, which will return the results after processing.

Given the layered distribution of eVTOLs, low-altitude tasks exhibit varying characteristics. eVTOLs in the low-speed layer excel at tasks requiring high precision and close-range operations, while eVTOLs in the high-speed layer, with their faster speeds and higher altitudes, are more suited to tasks requiring rapid responses and wide-coverage operations. Therefore, an optimal task offloading strategy should efficiently handle tasks at different layers simultaneously.

To describe eVTOL positions and their dynamic changes in the layered airspace, we establish a three-dimensional Cartesian coordinate system. In this system, BSs are located on the horizontal plane at $z=0$. 
For simplicity, we assume that all eVTOLs travel with horizontal velocities perpendicular to the x-axis. The dynamic movement of a given $eVTOL_i$, from time $t$ to $t+1$ is described by the following equations:
\begin{equation}
\begin{aligned}
x_i(t+1) &=x_{i}(t)\\
y_{i}(t+1) &=y_{i}(t)+v_{i,y}(t)\\
z_{i}(t+1)&=z_{i}(t)+v_{i,z}(t), 
\end{aligned}
\label{dsahfvytguasi}
\end{equation}

\noindent where $x_i(t+1)$, $y_i(t+1)$, and $z_i(t+1)$ denote the position of $eVTOL_i$ at time $t$, and $v_{i,y}$ and $v_{i,z}$ represent its horizontal and vertical velocity components, respectively.
At time $t$, the distance $d$ between $eVTOL_i$ and BS $N_j$ is given by:
\begin{equation}
\begin{aligned}
d_{i,N_j,t} = \sqrt{(x_{i}(t) - x_{N_j})^2 + (y_{i}(t) - y_{N_j})^2 + (z_{i}(t) - z_{N_j})^2}.
\end{aligned}
\label{fgdiugdtiu}
\end{equation}

In the layered AAM, eVTOL aviation undergoes three phases: takeoff, cruising, and landing.
During takeoff, the eVTOL’s vertical velocity is positive along the z-axis. When the eVTOL reaches its cruising layer, its vertical velocity becomes zero, and its horizontal velocity remains constant. As the eVTOL prepares to land, its vertical velocity becomes negative until it reaches the ground.

\subsection{Task Offloading Model}
Due to the limitations of onboard chips, an eVTOL must offload its tasks to a BS for computing and processing. The task offloading process consists of two main steps: First, the eVTOL establishes a wireless link with a nearby BS. Once the link is established, the eVTOL offloads the task data to the BS over this channel. 
Second, the offloaded task utilizes the computing, storage, and other resources provided by the slices deployed on the BS to complete the task.
It is important to note that the efficiency of task offloading varies when the eVTOL connects to different BS-Slice pairs. This variation arises from the differences in the amount of resources each slice accesses from the virtual resource pool.

At a given moment, suppose eVTOL $i$ releases a task request with the attributes $\left\{w_{i, ask},f_{i, ask},t_{i, ask},\mathcal{E}_{eVTOL_i}\right\}$. Here, $w_{i, ask}$ denotes the data volume of the task, $f_{i, ask}$ represents the required CPU cycles, $t_{i, ask}$ is the maximum tolerable delay, and the eVTOL dynamic information $\mathcal{E}_{eVTOL_i} =\{x_{i},y_{i},z_{i}, v_i\}$ includes the three-dimensional coordinates $(x_{i},y_{i},z_{i})$ and velocity $v_i$ of eVTOL $i$.
Assuming that the task is assigned to slice $q$, which contains resources ${V_{band_q},V_{beam_q},V_{comp_q}}$, representing the available bandwidth, beam alignment, and computing resources for slice $q$, respectively.
Moreover, due to the presence of multiple BSs as well as the additive Gaussian white channel \cite{9892688Zhang}, BS $N_j$ suffers from interference from other eVTOL-BS transmissions.
Thus, the achievable transmission rate from eVTOL $i$ to the BS $N$ via slice $q$ is:
\begin{equation}
\begin{aligned}
&r_{i,j,q}=V_{band_q}\cdot s_{band}\cdot \\ 
&\log_{2}(1+\frac{P_i|h_{i,N_j}|^2}{\sum_{oth\in N,oth \ne j}P_{oth}\cdot|h_{i,N_{oth}}|^2+\sigma^2}) \cdot \frac{G_{i,j}}{d_{i,N_j}^{2}}.
\end{aligned}
\label{rate}
\end{equation}

\noindent where $P_i$ is the transmission power of the eVTOL, $h_{i,N_j}$ is the channel gain from eVTOL $i$ to BS $N$, $\sigma^2$ is the thermal noise, and $G_{i,j}$ is the beamforming gain from eVTOL $i$ to BS $N_j$.
Additionally, the distance $d_{i,N_j}$ between eVTOL $i$ and BS $N_j$ is time-varying due to the mobility of the eVTOL.

We employ the beam gain model from \cite{1572261552770672512Sakaguchi}, which accounts for the relative positions of the eVTOL and BS, providing an accurate representation of beam gain variation. 
The beam gain \( G_{i,N_j} \) between eVTOL \( i \) and BS \( N_j \) is given as:
\begin{equation}
\begin{aligned}
&G_{i,N_j}=\frac{16}{6.76\varphi_{-3dB}}exp(-4\ln({2})\cdot\frac{\varphi_{i,N_j}^2}{\varphi_{-3dB}^2})\cdot V_{beam_q}\cdot s_{beam}.
\end{aligned}
\label{gain}
\end{equation}

\noindent Here, \( \varphi_{-3dB} \) is the beamwidth, and \( \varphi_{i,N_j} \), the azimuth angle of eVTOL \( i \) respect to \( N_j \), is:
\begin{equation}
\begin{aligned}
&\varphi_{i,N_j}=arc \cos \frac{z_{i,N_j}}{{d^2_{i,N_j}}}.
\end{aligned}
\label{tghdreuiu}
\end{equation}

In Eq.~\eqref{gain}, it is evident that $G_{i,N_j}$ is a function of $\varphi_{-3dB}$ and $\varphi_{i,N_j}$, where $\varphi_{i,N_j}$ continuously varies with the position of the eVTOL.
Substituting $G_{i,N_j}$ into Eq.~\eqref{rate}, and introducing the parameter  $\rho=\frac{P_i|h_{i,N_j}|^2}{\sum_{oth\in N,oth \ne j}P_{oth}\cdot|h_{i,N_{oth}}|^2+\sigma^2} $ for convenience, we obtain the derivative of $r_{i,j,q}$ with respect to $\varphi_{-3db}$:
\begin{equation}
\begin{aligned}
&\frac{dr_{i,j,q}}{d\varphi_{-3\text{dB}}} = \frac{S \cdot Q \cdot \left( 8 \ln(2) \varphi^2 \varphi_{-3\text{dB}}^{-4} - \varphi_{-3\text{dB}}^{-2} \right)}{ \ln(2) \cdot \left( 1 + S \varphi_{-3\text{dB}}^{-1} \cdot  Q  \right)}.
\end{aligned}
\label{akddbdkde}
\end{equation}

\noindent Here, $Q=-\exp(4 \ln(2) \frac{\varphi^2}{\varphi_{-3\text{dB}}^2})$ and $S=\frac{16\rho}{6.76 d_{i,N_j}^2}$.
By setting the derivative to zero, we can determine the maximum beamforming gain, $G_{i,N_j}$, corresponding to the maximum achievable rate:
\begin{equation}
\begin{aligned}
&G^{max}_{i,N_j}=\frac{16}{6.76 \cdot \varphi_{i,N_j} \sqrt{8\ln(2)\cdot e}}.
\end{aligned}
\label{beirltujgirte}
\end{equation}

\noindent For simplicity, this paper assumes that the beam gain between eVTOLs and BSs equals the maximum beam gain. The transmission delay for task offloading can thus be described as:
\begin{equation}
\begin{aligned}
&t_{i,N_j,q,tran}=\frac{w_{i,ask}}{r_{i,N_j,q}}.
\end{aligned}
\label{vwrtiugfhweroug}
\end{equation}

\noindent Once the task is offloaded to the BS, it is processed by the computing resources allocated to the slice. Thus, the computational delay is:
\begin{equation}
\begin{aligned}
t_{i,N_j,q,comp}=\frac{f_{i,ask}}{V_{comp_q}\cdot s_{comp}}.
\end{aligned}
\label{fewriufero}
\end{equation}

Since the BS transmission power is much higher than that of eVTOL, and the size of the post-processed task is smaller, the downlink feedback delay is typically negligible. Therefore, the total task offloading delay is the sum of the transmission and computational delays,
\begin{equation}
\begin{aligned}
t_{i,N_j,q}=t_{i,N_j,q,tran}+t_{i,N_j,q,comp}.
\end{aligned}
\label{fdsguegeiioiuf}
\end{equation}

\subsection{Low-altitude Slice Model}
Due to the limited onboard resources of eVTOLs, they often offload tasks to a nearby BS that leverages the dynamically allocated slice for offloading processing. In our proposed scenario, each BS possesses multiple types of low-altitude slices, with each slice acquiring resources from the virtual resource pool abstracted by the resource mapping layer.

Let the set of BSs be denoted as \( N = \{ N_1, N_2, \dots, N_m \} \), where \( m \) represents the total number of BSs.
The set of slices is defined as \( S = \{ S_1, S_2, \dots, S_n \} \), where \( n \) represents the total number of low-altitude slices.

When an eVTOL initiates a resource request from a specific slice, it must first access to the BS that hosts the slice. If the request is accepted, the eVTOL establishes a wireless connection with the selected BS \( N_j \), which provides the slice \( S_q \), thereby granting the required resources.

To better match the needs of offloaded tasks, the proposed slice scheme customizes slice attributes based on task requirements, rather than relying on predefined slice types from 5G/6G systems. Consequently, each slice can dynamically extract varying amounts of resources from the resource mapping layer to serve the tasks assigned to it.

The total resource pool from the mapping layer is represented as \( s = \{ s_{band}, s_{beam}, s_{comp} \} \), where $s_{band}$, $s_{beam}$, and $s_{comp}$ denote the total bandwidth, antenna beam, and computing resources available in the scenario, respectively. At each time slot, each slice selects a portion of these resources to form the slice's resource allocation for that moment.

Let \( \{ V_{band_q}(t), V_{beam_q}(t), V_{comp_q}(t) \} \) represent the proportions of resources allocated to slice \( q \) from the resource pool at time $t$, where each value is expressed as a percentage. Therefore, the actual resources obtained by slice \( q \) at time $t$ are:
\begin{equation}
\begin{aligned}
s_q(t) &= \{ V_{band_q}(t) \times s_{band}, V_{beam_q}(t) \times s_{beam}, \\ & V_{comp_q}(t) \times s_{comp} \}. 
\end{aligned}
\label{usgsgkeru}
\end{equation}

For a low-altitude slice, the amount of bandwidth resources determines the transmission capacity of air-ground communication. 
Antenna beam resources reflect the alignment of the antenna used for tracking an eVTOL’s flight. 
The more beam resources available, the stronger the antenna’s ability to track high-mobility eVTOLs, which improves channel gain and communication quality. 
Computing resources are used to process offloaded tasks, and a greater amount of computing power leads to faster task processing.
To better describe dynamic in slice resources during a task cycle, we introduce a slice lifecycle that consists of three phases: "initialization," "scaling," and "disposal."

\emph{Initialization}: When an eVTOL generates a new task, the resource amounts for each slice are initialized based on the task's requirements. Typically, the initial resource allocation is set to a preset value based on historical records.

\emph{Scaling}: During task processing, the eVTOL's resource demands may change dynamically over time. To better serve real-time tasks, each slice adjusts its resource allocation accordingly. For instance, if a task requires more bandwidth, the slice can increase the proportion of bandwidth resources allocated to it. Conversely, when the tasks require relatively less bandwidth, the slice will reduce the proportion of bandwidth resources and release the excess resources to other slices.

\emph{Disposal}: Once all tasks assigned to a slice are completed, the slice will release its resources back to the virtual resource pool of the mapping layer. This avoids resource waste and allows the resources to be reallocated to other slices, improving global resource utilization.

%

\subsection{System Objective}

Based on the system configurations, we formulate a slice resource optimization problem that aims to maximize user satisfaction while minimizing resource consumption. This problem considers the heterogeneous resources, dynamic task offloading, and the specific mobility of eVTOLs.
First of all, to evaluate task offloading performance, we propose two key metrics for network slice optimization: \emph{user satisfaction} and \emph{resource consumption}.

\emph{User satisfaction} reflects the extent to which a slice meets the task requirements. When eVTOL \( eVTOL_i \) accesses slice \( S_q \), the user satisfaction can be expressed as:
\begin{equation}
\begin{aligned}
Sat_{i,q}=\frac{1}{1+e^{-\eta(t_{i,ask}-t_{i,q})}},
\end{aligned}
\label{brtdugyhwei}
\end{equation}

\noindent where \( \eta \) is a weight that adjusts the satisfaction range. If \( t_{i,q} \) exceeds \( t_{\text{ask}} \) (the maximum acceptable delay), the basic task requirements are met, and satisfaction \( \text{Sat}_{i,q} \) increases. Conversely, if \( t_{i,q} < t_{\text{ask}} \), satisfaction decreases.
 
Since a slice \( S_q \) can accommodate multiple eVTOLs, a typical \emph{average user satisfaction} of the slice is given by:
\begin{equation}
\begin{aligned}
Sat_q=\sum_{i\in S_q}{Sat_{i,q}}.
\end{aligned}
\label{gwetruy}
\end{equation}

\noindent Here, higher satisfaction indicates better task completion for the slice. However, if tasks are not completed on time, it can disrupt subsequent eVTOL operations, leading to lower aviation efficiency or even accidents. To account for this, we impose a penalty on slices when they fail to complete tasks on time. This penalty is referred to as the \emph{violation cost} and is defined as:
\begin{equation}
\begin{aligned}
C_{q,v}=\omega_v \sum_{i\in S_q} I\left\{t_{i,N_j,q}>t_{i,ask}\right\},
\end{aligned}
\label{dfugsueig}
\end{equation}

\noindent where \( I\{x\} \) is an indicator function that equals $1$ when \( x \) is true, otherwise $0$. 
\( \omega_v \) represents the cost weight when the task offloading time $t_{i,N_j,q}$ exceeds the maximum allowable delay $t_{i,ask}$.


\emph{Resource consumption} refers to the resources allocated to a slice, including bandwidth, beam alignment, and computing power. Let \( C_{q,o} \) represent the operational resource cost of slice \( S_q \), which is expressed as:

\begin{equation}
\begin{aligned}
C_{q,o}&=\omega_{band}V_{band_q}s_{band}+\omega_{beam}V_{beam_q}s_{beam_q}\\
&+\omega_{comp}V_{comp_q}s_{comp},
\end{aligned}
\label{ghydhduihih}
\end{equation}

\noindent where \( \omega_{\text{band}}\), \(\omega_{\text{beam}}\), and \(\omega_{\text{comp}} \) represent the cost weights for acquiring bandwidth, beam alignment, and computational resources, respectively. A higher allocation of bandwidth, beam alignment, and computational resources results in greater resource consumption.

By considering both the \emph{violation cost} and the \emph{operational cost}, the total cost of a slice is the sum of these two components:

\begin{equation}
\begin{aligned}
C_q=C_{q,v}+C_{q,o}.
\end{aligned}
\label{ytriuro}
\end{equation}

The total cost of a slice reflects not only the resource consumption from the resource pool but also the slice's ability to complete the assigned tasks on time. It serves as a unified quantitative metric for evaluating resource utilization efficiency and task execution quality. This enables the optimization of slice resource allocation and eVTOL task offloading scheduling.

\subsection{Optimization Model}
To optimize both user satisfaction and the total cost of slices comprehensively, the overall resource allocation problem can be formulated as:

\begin{subequations}
\begin{align}
\textbf{P1:}~ &~\max \sum_{q \in \left\{1,2,...n\right\}}({\alpha Sat_{q} - \beta C_{q}})\\
\text{s.t.} 
&~C1: \ V_{band_q}, V_{beam_q},V_{comp_q}\in[0,1],\\
&~C2:\sum_{q\in \left\{1,2,...n\right\}}V_{band_{q}}\leq1,\\
&~C3:\sum_{q\in \left\{1,2,...n\right\}}V_{beam_{q}}\leq1,\\
&~C4:\sum_{q\in \left\{1,2,...n\right\}}V_{comp_{q}}\leq1,\\
&~C5: \{x_i(t),y_i(t),z_i(t)\} \in \mathcal{R}_{cor}, \\
&~C6: v_i(t)=\{v_{low}, v_{high}\}, \forall i \in N.
\end{align}
\end{subequations}
\label{ugvuadsavyi}

\noindent Here, constraint $C1$ specifies the resource proportion allocated to each slice. Constraints $C2$, $C3$, and $C4$ ensure that the total allocation of each resource type does not exceed the total available resources in the virtual resource pool. Constraint $C5$ indicates that eVTOLs must follow the prescribed layered airline $\mathcal{R}{cor}$. Lastly, $C6$ defines the velocity of each eVTOL, where $v_i(t)$ is set to $v{low}$ for low-speed layers, and $v_{high}$ for high-speed layers.

The problem, characterized by non-convex objective functions (user satisfaction and resource consumption), is inherently a non-convex optimization issue. Furthermore, the complexity of the optimization is influenced by the eVTOL mobility, which results in highly time-varying resource demands for tasks. This makes the optimization of resource allocation and slice lifecycle management particularly challenging.

\begin{figure}[h]
\centering
     \includegraphics[width=.4\textwidth]{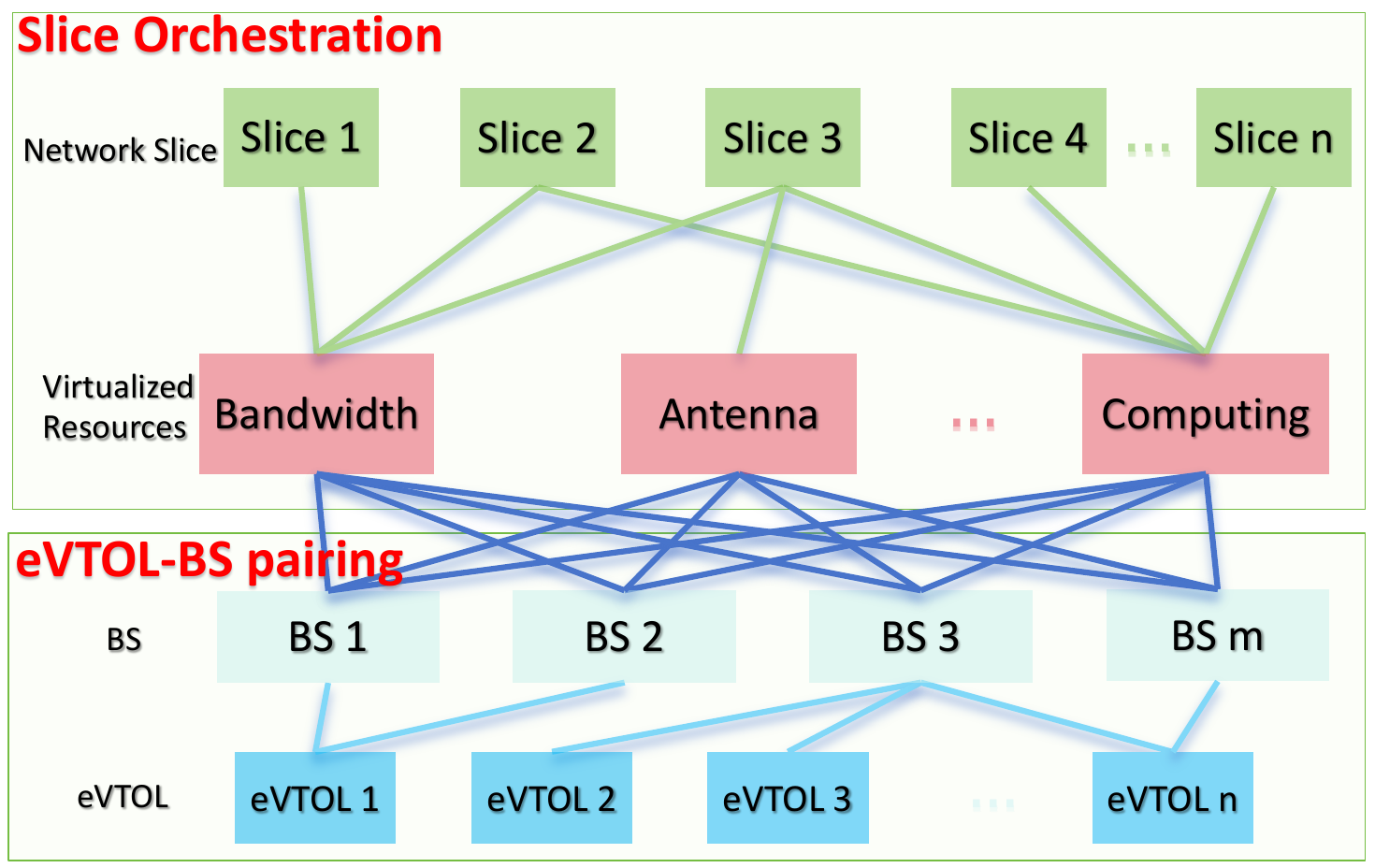} 
     \caption{Demonstration of Intelligent Slicing Orchestration.} 
\label{slice_match_scheme}
\end{figure}

As shown in Fig.~\ref{slice_match_scheme}, we divide the optimization process into two subproblems. The first subproblem focuses on determining the optimal eVTOL-BS access pairing, while the second addresses the optimal resource allocation for slice orchestration, balancing user satisfaction and resource consumption.

We begin by proposing a task priority matching algorithm, which aims to pair dynamic eVTOLs with the optimal access BS. The algorithm takes into account the task completion time requirements and the relative positions of eVTOLs and BSs, ensuring efficient task offloading.

Building on the eVTOL-BS access pairing, we then introduce a resource allocation strategy based on Multi-Agent Reinforcement Learning (MARL) to manage dynamic slice resource orchestration throughout the slice lifecycle. The MARL algorithm dynamically adjusts the resource allocation ratio of each slice in real time, enabling the expansion or contraction of slice resources based on task offloading needs.

\section{Model Solution}
\begin{figure}[h]
\centering
     \includegraphics[width=.48\textwidth]{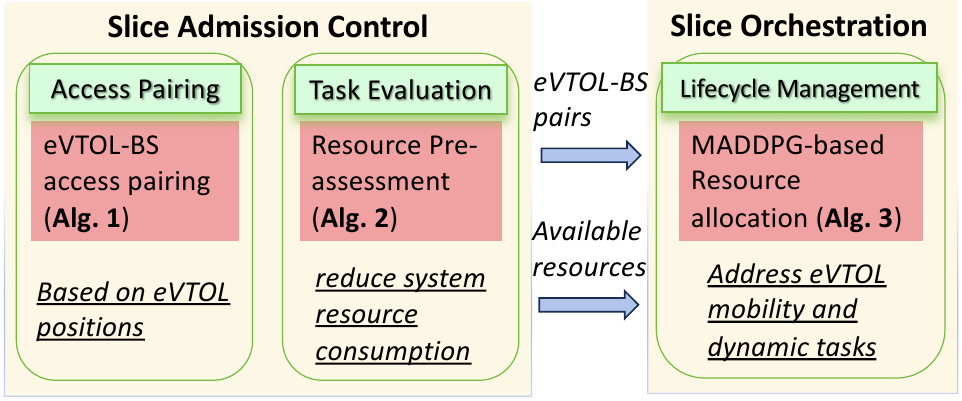} 
\caption{Relations of the proposed three algorithms.} 
\label{Solution_Scheduling}
\end{figure}

This section proposes three algorithms to solve $P1$ using the VNSM framework. The solution involves two main components: the slice admission control module and the slice orchestration module of the VNSM. The relationships between the algorithms, modules, and their corresponding functions are illustrated in Fig.~\ref{Solution_Scheduling}.

As shown in Fig.\ref{Solution_Scheduling}, the slice admission control module includes two algorithms: the access pairing algorithm and the resource pre-assessment algorithm. During the eVTOL access stage, the access pairing algorithm (Alg.~\ref{Ag1}) determines the optimal eVTOL-BS pairing based on the distances between eVTOLs and BSs. In the task evaluation stage, the pre-assessment algorithm (Alg.~\ref{Ag2}) allocates the total available resources based on the task requirements, aiming to minimize resource wastage.

Through Alg.~\ref{Ag1} and Alg.~\ref{Ag2}, the slice admission control module generates the eVTOL-BS pairing and defines the total available resources for slice orchestration. Then, in the lifecycle management stage, resource allocation is handled using a MADDPG-based approach (Alg.~\ref{Ag3}), considering the mobility of eVTOLs and the dynamic nature of task demands.

Thus, the entire process starts with eVTOL-BS access pairing, progresses through resource pre-assessment to reserve available resources, and concludes with resource allocation using the MADDPG-based approach. This ensures efficient task offloading and optimal resource utilization.

\subsection{eVTOL-BS Access Pairing Scheme}
The mobility of eVTOLs result in a constantly changing distance between an eVTOL and its access BS during flight. This variable distance influences the channel conditions between the eVTOL and the BS, which in turn affects the choice of access BS and the corresponding slice for task offloading. Additionally, since the resources allocated to different slices vary, the channel conditions between the eVTOL and the BS hosting a specific slice will change as the eVTOL connects to different slices. Therefore, it is essential to consider the optimal matching between the eVTOL, BS, and the hosting slice in a comprehensive manner.

Hereafter, This section introduces the eVTOL-BS Access Pairing algorithm (eBAP) for task offloading, which aims to maximize user satisfaction. The matching priority between eVTOL \( i \), BS \( N_j \), and slice \( S_q \) is defined as follows:
\begin{equation}
\begin{aligned}
&\omega_{i,N_j,q}=\frac{\gamma \frac{t_{i,ask}}{t_{i,N_j,q}}+(1-\gamma)\frac{1}{d_{i,N_j}}}{\sum^{m}_{i=1}{\gamma \frac{t_{i,ask}}{t_{i,N_j,q}}+(1-\gamma)\frac{1}{d_{i,N_j}}}}.
\end{aligned}
\label{cdrafvkafb}
\end{equation}

\noindent In this equation, \( \omega_{i,N_j,q} \) represents the priority for resource allocation, and \( \gamma \) is a weight parameter that controls the balance between the two terms in the priority calculation. The first term, \( \frac{t_{i,ask}}{t_{i,N_j,q}} \), reflects the task-solving efficiency. A higher value indicates that the resources provided by BS \( N_j \) and slice \( S_q \) result in a shorter task completion time \( t_{i,N_j,q} \) for eVTOL $i$, which leads to more efficient task offloading. Here, $t_{i,ask}$ is the maximum allowed offloading time for eVTOL $i$’s task.

The second term, \( \frac{1}{d_{i,N_j}} \), is introduced to account for the distance between eVTOL $i$ and BS $N_j$. This term ensures that, when multiple BS-slice pairs provide similar resources, the algorithm prioritizes the closest BS \( N_j \) to maximize wireless transmission performance.

After calculating the matching priority \( \omega_{i,N_j,q} \) for each eVTOL $i$ and its corresponding BS $N_j$-slice $S_q$ pair, eVTOL $i$ will attempt to connect to the BS-slice pairs in descending order of priority. A higher \( \omega_{i,N_j,q} \) corresponds to a higher probability of access. If the current eVTOL-BS-slice matching attempt fails (due to task demands exceeding the resource capacity of the BS), the process will immediately move to the next BS-slice pair with a lower priority. This continues until eVTOL $i$ successfully connects to a BS-slice pair. The matching process concludes when all eVTOLs are successfully paired with BS-slice pairs.
The specific steps of the eBAP matching algorithm are as Alg.~\ref{Ag1}:

\begin{algorithm} 
	\caption{eVTOL-BS Access Pairing algorithm} \label{Ag1}
	
    \textbf{Input}: distance $d_{i,N_j}$ between eVTOL and BS, the maximum tolerable delay $t_{i,ask}$
    
    \textbf{Output}: eVTOL-BS-Slice pairs
    
    Calculate all available $\omega_{i,j,q}$ between $eVTOL_i$, $N_j$, and $S_q$ by Eq.~(\ref{cdrafvkafb});

    \For{$i =0,1,2,\dots,m$} 
    {
		Select the BS $N_j$ and slice $S_q$ corresponding to the maximum $\omega_{max,i,N_j,q}$;
        
   
        \If{$N_j$ or $S_q$ has reached its capacity limit for access} {
            Select the BS with the highest priority among the remaining BSs for access pairing;
        }
        \Else{
            $eVTOL_i$ access BS $N_j$ with slice $S_q$;
        }		
    }				
\end{algorithm}

\subsection{Resource Pre-assessment}

The resource demands generated by eVTOLs can vary significantly. Under high task loads, it is crucial to fully utilize network resources to meet task requirements. However, under low task loads, excessive resource allocation may lead to unnecessary waste. To address this, this section introduces an efficient resource pre-assessment algorithm that evaluates tasks before the actual resource allocation.

The proposed resource pre-assessment algorithm incrementally explores different levels of total resource allocation. By gradually increasing the allocated resource volume, the algorithm identifies the minimal amount of resources required to meet task demands. Note that this process determines the total resource demand before the slice orchestration, thereby preventing unnecessary resource occupation and waste in the slice operations.

When an eVTOL generates a task, a set of task configuration parameters is extracted, denoted as $\left\{w_{avr}, f_{avr}, t_{avr}, \varepsilon_{i}\right\}$, where: $w_{avr}, f_{avr}, t_{avr}$ represent the average data volume, required computing cycles, and allowed delay requirements of all tasks in the current network, respectively. 
Meanwhile, $\varepsilon_{i}$ represents the attributes of eVTOL $i$, which includes its three-dimensional coordinates and velocity. The spatial coordinates of eVTOLs are retained individually to preserve their positional information, rather than averaging them.

Next, the averaged tasks are reallocated to all eVTOLs. A traversal mechanism is introduced to manage resource allocation levels, denoted as $l \, (l = 1, 2, \dots, l_{max})$. Here, $l_{max}$ represents the maximum resource allocation level. The allocation begins at the minimum level and increases incrementally. For each level $l$, the total available resources are calculated as $l \times \left\{s_{band}, s_{beam}, s_{comp}\right\}$, with the total resources increasing as the allocation level rises.

At each resource allocation level, available resources are used to meet task demands. If the current resources cannot satisfy the offloading delay requirements for all tasks, the resource allocation level is increased. Resources are then reallocated, and this process continues until the delay requirements for all eVTOL tasks are met. The current resource allocation level is considered the minimal allocation for the task. The process concludes once the optimal level is determined. Any additional resources beyond this level would result in waste.
It is important to note that if the traversal reaches $l = l_{max}$ and the task requirements are still unmet, $l$ is still selected to the maximum available level $l_{max}$. Moreover, we present the process of resource pre-assessment algorithm in Alg.~\ref{Ag2}.

\begin{algorithm}[!h]
    \caption{Resource Pre-assessment algorithm} \label{Ag2}
    
    \textbf{Input}: total amount of resources $\left\{s_{band},s_{beam},s_{comp}\right\}$,
    tasks $\left\{w_{i},f_{i},t_{i}\right\}$ generated by eVTOL $i$ 
    
    \textbf{Output}: resource allocation level for current stage $l$
    
    Calculate the average demands of all tasks $\left\{w^{ave},f^{ave},t^{ave}\right\}$;

    Assign the averaged demands to all eVTOLs;
    
    \For{$l =0,1,2,\dots i_{max}$} 
    {
        Pre-set the available resource $R=l \times \left\{s_{band},s_{beam},s_{comp}\right\}$;

        Calculate the task delay $t_{i,j,q}$ based on the pre-set resource;

        \If{$t_{i,j,q} > t^{ave}$ }
        {   
            Increase allocation level $l+=1$;
        }
        \Else
        {    
            break;
        }
        
    }

\end{algorithm}

\subsection{MADDPG-based Slice Orchestration}
eVTOLs continuously operate in three-dimensional space, meaning that their positions relative to the BSs change over time. Furthermore, the tasks generated by the eVTOLs also vary, leading to dynamic shifts in the relative distance between the eVTOL and the BS. This, in turn, impacts the access pairing between the eVTOL and the service slice. Consequently, the optimization problem becomes high-dimensional and challenging, as it is difficult to find a globally optimal solution. In general, such problems, which involve interactions with the environment, can be addressed using Reinforcement Learning (RL).

However, single-agent RL faces limitations. An agent can only make decisions based on its local information and interactions with the environment, making it difficult to fully consider the broader tasks and resource interactions among multiple agents from the system perspective.
Additionally, in complex scenarios involving multiple eVTOLs, BSs, and slices, single-agent RL models struggle to navigate the intricate interactions, often leading to local optima.
To overcome these challenges, this paper adopts a Multi-Agent RL framework. In this framework, each slice is modeled as an independent agent capable of autonomous decision-making. Through collaboration, these agents aim to optimize the global objective function, improving the efficiency and performance of system resource allocation. 

However, a key challenge lies in the continuous action and state spaces of resource allocation. Traditional RL methods are effective for problems with discrete state spaces but struggle with continuous ones \cite{9410457Chongwen}. Fortunately, the Deep Deterministic Policy Gradient (DDPG) method is well-suited for continuous variable optimization. This method has been widely applied in various communication resource management problems \cite{{9410457Chongwen},{10669355kai}}.
Specifically, each slice is modeled as an independent agent, making decisions autonomously. Through collaborative optimization, the agents collectively maximize the global objective. Therefore, we utilize the Multi-Agent DDPG (MADDPG) algorithm, which extends DDPG to multi-agent scenarios. MADDPG is a distributed Multi-Agent RL framework, where each agent employs a DDPG model to make decisions based on local observations. By using a well-designed reward function, the agents can exhibit both cooperative and competitive behaviors, ultimately optimizing system performance \cite{cui2023multi}.

In reinforcement learning, problems are often modeled as a Markov Decision Process (MDP). An MDP is defined by a tuple $(S, A, P, R, \gamma)$, where:
$S$ represents the state space, describing the system's state at any given moment.
$A$ is the action space, which defines the set of actions an agent can take in each state.
$P$ is the state transition probability, specifying the likelihood of transitioning from one state to another.
$R$ is the reward function, which defines the immediate reward an agent receives after taking an action in a given state.
$\gamma \in [0,1]$ is the discount factor that balances immediate rewards against long-term benefits.
Next, we detail the components of the MDP for the low-altitude slice system.

\textbf{State:} The state reflects the current resource allocation, average user satisfaction, and system costs for a given slice under a specific action (allocation strategy). A single agent's partial observation is defined as a triplet $[V_q, C_q, Sat_q]$, where $V_q$ represents the allocated resource ratio for slice. $C_q$ represents the cost of the slice, and $Sat_q$ denotes the slice’s average user satisfaction. Specifically, $V_q$ includes the proportions of bandwidth, beam alignment, and computing resources in the resource pool, i.e., $V_{band_q}, V_{beam_q}, V_{comp_q}$. The cost $C_q$ and user satisfaction $Sat_q$ are calculated using the formulas in Eq.~(\ref{ytriuro}) and Eq.~\ref{gwetruy}, respectively.
The overall system state is the aggregation of all agents' partial observations. For instance, at time $t$, the state of the system is represented as:
\begin{equation}
\begin{aligned}
s(t)=\left\{\left\{V_q(t),C_{q}(t),Sat_{q}(t)\right\},q\in \left\{1,2,\dots,n\right\}\right\}
\end{aligned}
\label{vsuyvseuirrgua}
\end{equation}

In this case, \(\left\{V_q(t), C_q(t), Sat_q(t)\right\}\) is considered the partial observation \(o_q\) of agent \(q\). During the decision-making process, each agent selects actions independently based on its local observation. Although each agent has its own local observations, they are not entirely independent. Agents can leverage the observations and action histories of all agents to train a centralized critic network, aligning their actions with the global objective.

\textbf{Action:} The action represents the behavior of adjusting the resource proportions of a slice, i.e., increasing or decreasing the proportion of resources based on the slice’s original allocation. For example, an action of $-0.2$ means decreasing the resource allocation by $20\%$. As the orchestration involves three types of resources (bandwidth, beam alignment, and computing), the action is a vector representing the adjustment proportions for each resource. This is given by:
\begin{equation}
\begin{aligned}
a(t)=\left\{\left\{a_{band_q}(t),a_{beam_q}(t),a_{comp_q}(t)\right\},q \in \left\{1,2,\dots,n\right\}\right\}
\end{aligned}
\label{fdsguegeiiodiuf}
\end{equation}

\noindent Here, $a_{band_q}(t), a_{beam_q}(t), a_{comp_q}(t)$ represent the adjustment proportions for bandwidth, antenna beam, and computational resources for slice $S_q$ at time $t$, respectively. These values are continuous variables ranging from -1 to 1.

\textbf{Reward:} An agent’s reward is designed to reflect its satisfaction in handling offloaded tasks and the proportion of resource consumption. If an agent occupies too much of the available resources, it negatively impacts the overall system's task-handling efficiency, as other agents will have insufficient resources to handle their tasks. This situation should be avoided. To prevent this, each agent’s reward is defined as the weighted sum of the system's total cost $C_q$ and user satisfaction $Sat_q$ for the slice, given by:
\begin{equation}
\begin{aligned}
r^{t}_{q}=\omega_1\cdot (Sat_q-0.5)-\omega_2\cdot C^{t}_{q}
\end{aligned}
\label{argaidrghubdv}
\end{equation}

\noindent Here, $\omega_1$ and $\omega_2$ are the weights for user satisfaction and the system’s total cost, respectively. In this design, $(Sat_q - 0.5)$ serves as the first term of the reward function. This term is positive when the user’s requirements are largely satisfied and negative otherwise, helping to prevent scenarios where a slice receives inadequate resources to complete its tasks effectively.
 
Under this framework, agents explore the state and action spaces of the MDP by interacting with the environment, aiming to discover a policy $\pi(a|s)$ that maximizes their rewards. In a multi-agent context, each agent operates based on its partial observations of the state and action spaces. Agents influence each other through the environment and reward functions.

Based on the MADDPG algorithm, slices can adjust their resource allocation in real time based on task demands. Initially, each slice’s resources are set according to a predetermined proportion. As the task progresses, each slice, as an agent, adjusts its resource proportions dynamically to optimize rewards. Once the task concludes, the slice’s resources are returned to the resource pool for future use by other slices.

\begin{figure}[h]
\centering
     \includegraphics[width=.48\textwidth]{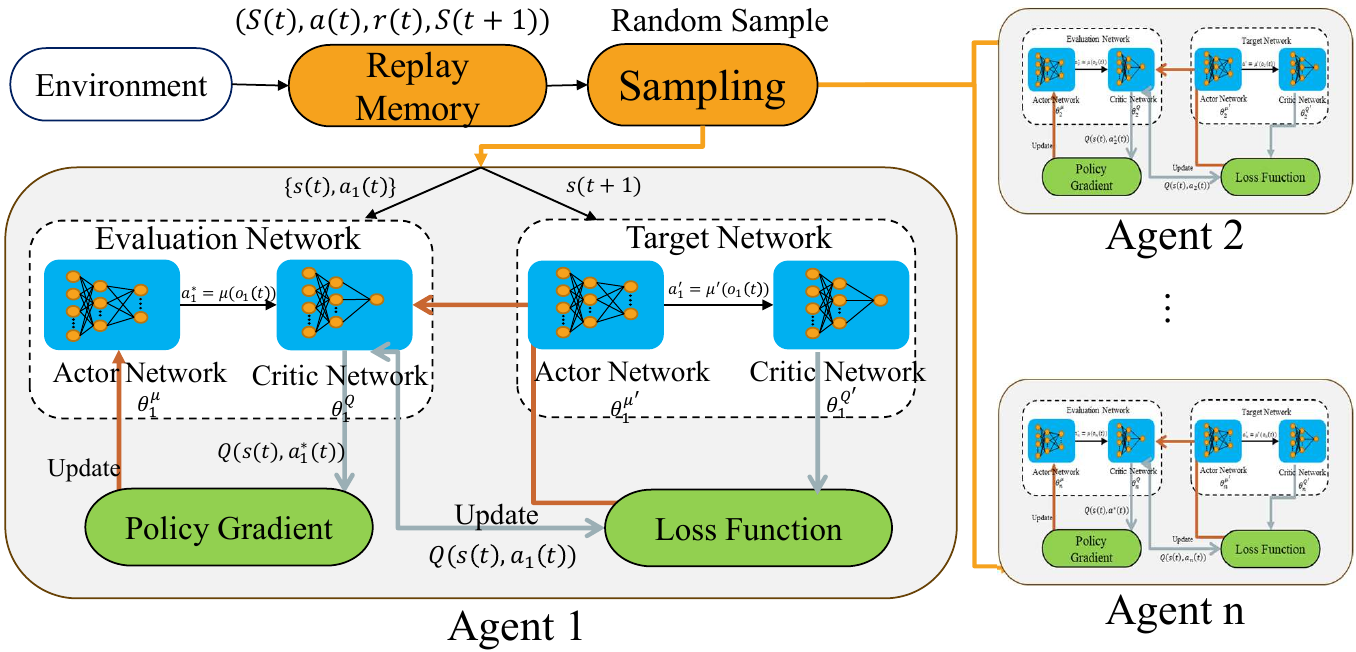} 
\caption{MADDPG-based Resource Allocation Framework.} 
\label{DDPG framework}
\end{figure}



The MADDPG framework is shown in Fig.~\ref{DDPG framework}, consisting of an external environment and $n$ agents. Each agent within the framework deploys a DDPG structure to facilitate its learning and decision-making processes. The DDPG method comprises two primary components: the evaluation network and the target network. Each component includes both an actor network and a critic network. The actor network generates actions based on the current policy, while the critic network evaluates the quality of the actions taken by the actor network.

In the context of the MADDPG framework, which involves multiple agents, the parameters of the actor and critic networks in the evaluation network for agent $m$ are denoted as $\theta^{\mu}_{m}$ and $\theta^{Q}_{m}$, respectively. Similarly, the parameters of the actor and critic networks in the target network are represented as $\theta_{m}^{\mu^{'}}$ and $\theta_{m}^{Q^{'}}$. Here, the subscript $\mu$ refers to the actor network parameters, $Q$ indicates the critic network parameters, and the superscript $m$ identifies the specific agent.

Agents gather experiences through interactions with the environment, which include states $S(t)$, actions $a(t)$, rewards $r(t)$, and subsequent states $S(t+1)$. During training, agents continuously interact with the environment and acquire additional experiences. These experiences are stored in a replay memory. When it is time to update the network parameters, experiences are randomly sampled from this memory for the update process.

When updating the network for agent $m$, the evaluation network is updated based on the state-action pairs $\left\{s(t),a(t)\right\}$. In contrast, the target network is updated using the subsequent states $s(t+1)$ derived from the experiences in the replay memory. The agents synchronize updates to their evaluation and target networks using the following equations \cite{cui2023multi}.
The loss function for the critic network in the evaluation network is defined as:
\begin{equation}
\begin{aligned}
L(\theta_{m}^{q})=\frac{1}{K}\sum_{i=1}^{K}(r_{m}^{i}+\gamma Q_{m}^{'}(s_{m}^{i+1},\mu_{m}^{'}(s_{m}^{i+1})\mid \theta_{m}^{Q^{'}})
-\\Q_{m}(s_{m}^{i},a_{m}^{i}\mid \theta_{m}^{q})
\end{aligned}
\label{criticupdate}
\end{equation}
where $\gamma$ is the discount factor, and $\mu_{m}^{'}(s_{m}^{i+1})$ represents the output action from the target actor network, with $s_{m}^{i+1}$ as the input.
Thus, the policy gradient for the actor network in the evaluation network is given by:
\begin{equation}
\begin{aligned}
\bigtriangledown _{\theta _m^\mu}J(\theta _m^\mu)\approx\frac{1}{K}\sum_{i=1}^{K}[\bigtriangledown _{\theta _m^\mu}\mu_m(s_m)\mid _{s_m=s_m^i}\\
\bigtriangledown_{a_m}Q(s_m,a_m\mid 
\theta _m^Q)\mid _{s=s_m^i,a_m=\mu_m(s_m^i)}]
\end{aligned}
\label{actorupdate}
\end{equation}
Finally, the specific procedure for the MADDPG algorithm is presented in Alg.~\ref{Ag3}.

\begin{algorithm} 
	\caption{MADDPG-based Resource Allocation} \label{Ag3}
	
    \textbf{Input}: dynamics of eVTOLs, eVTOL-BS parings, total available resource
    
    \textbf{Output}: resource allocation policy $\pi(a|s)$

	\textbf{Initialization}: experience replay memory; actor network ${\theta_m ^\mu }$; target actor network ${\theta_m ^{\mu'}}$, critic network ${\theta_m ^Q}$, target critic network ${\theta_m ^{Q'}}$ for agent $m$
    \\
	\For{$episode$ =0,1,2,...max episode} 
	{
		Observe initial state $s(0)$\\
		\For {$step$ = 0,1,2,...max step} {
        Each agent get observation and generate action $a(t+1) = \pi(\theta_m ^\mu | s(t), a(t))$ from actor network

        Observe new state ${s(t + 1)}$ and reward $r(t)$

        Save $\left\{{s(t)},{a(t)},{r(t)},{s(t + 1)}\right\}$ in experience replay memory
   
    \If{experience replay memory is full} {
                Take a random sample from the memory
    
                 Each agent separately update loss function, for agent $m$, it can be updated according to (\ref{criticupdate}) and (\ref{actorupdate})

                 Each agent separately update the target network, for agent $m$, it can be updated according to:
                 
                 $\begin{gathered}
  {\theta_m ^{Q'}} = \tau {\theta_m ^Q} + (1 - \tau ){\theta_m ^{Q'}} \hfill \\
  {\theta_m ^{\mu '}} = \tau {\theta_m ^\mu } + (1 - \tau ){\theta_m ^{\mu '}} \hfill \\ 
\end{gathered}$\\
			}
		}
	}
\end{algorithm}

\section{Performance Evaluation}
This section conducts numerical simulations to assess the performance of the proposed low-altitude slice scheme in the AAM scenario. The algorithms under consideration include access pairing and resource pre-assessment for slice admission control, as well as DDPG-based resource allocation for slice orchestration.

\begin{figure}[h]
\centering
     \includegraphics[width=.4\textwidth]{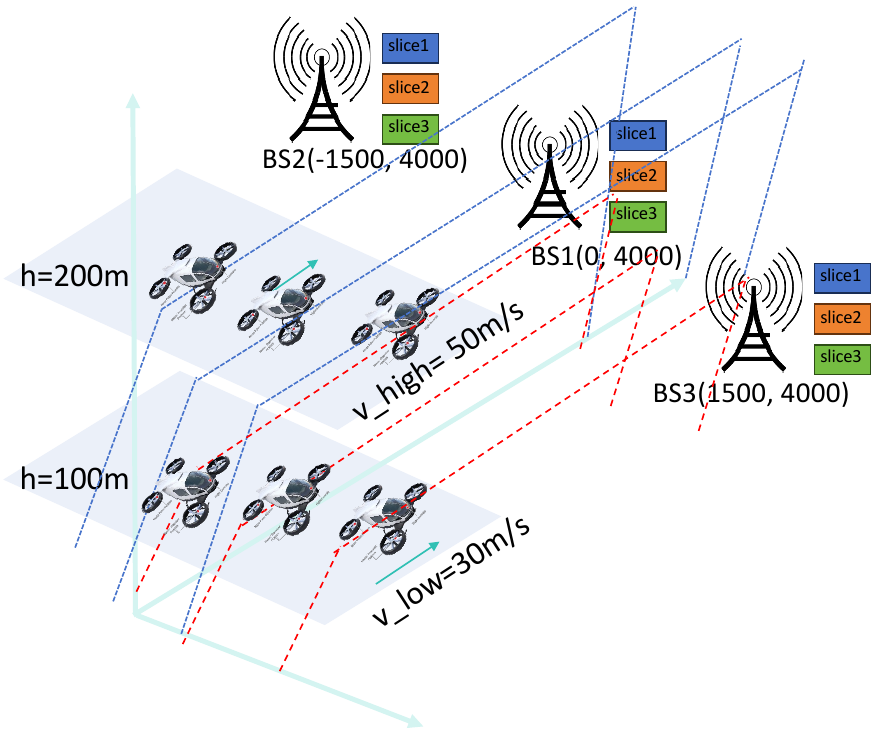} 
\caption{Simulated scene configuration.} 
\label{sim_scenario}
\end{figure}

First, we present the simulation configurations, as illustrated in Fig.~\ref{sim_scenario}. The AAM scenario consists of three BSs positioned on the ground and six eVTOLs. The eVTOLs are divided into two layers: three in the high-speed layer and three in the low-speed layer. The BSs are located at coordinates $(-1500, 4000, 0)$, $(0, 4000, 0)$, and $(1500, 4000, 0)$, respectively.
The low-speed layer is positioned 100 meters above the ground, with a constant velocity of $30 m/s$. The high-speed layer is located 200 meters above the ground, with a prescribed velocity of $50 m/s$.
We implement the proposed algorithms using PyTorch. A summary of the key simulation parameters is provided in Tab.~\ref{Simulation parameters}.

\begin{table}[h]
\centering
\caption{Simulation parameters}
\begin{tabular}{|c|c|}
\hline
Number of eVTOLs & 6  \\
\hline
Number of BSs & 3 \\
\hline
Maximum simultaneous access for a BS $m$ & 3 \\
\hline
Maximum simultaneous access for a slice $n$ & 2 \\
\hline
Total amount of bandwidth resources  $s_{band}$ & 100 MBps \\
\hline
Total amount of computing resource $s_{comp}$ & 100 GFLOPS\\
\hline
Maximum resource allocation level $l_{max}$ & 10 \\
\hline
Weight that adjusts the satisfaction range $\eta$ & 0.1 \\
\hline
Unit cost of the QoS violation $\omega_v$ & 2 \\
\hline
Learning rate for training critic network in DDPG & 0.00001 \\ \hline
Learning rate for training actor network in DDPG & 0.00002 \\ \hline
Soft replacement in DDPG & 0.01 \\ \hline
Buffer size for experience replay & 15000 \\ \hline
Numbers of episodes in DDPG & 800 \\ \hline
Numbers of steps of each episode in DDPG & 300 \\ \hline


\end{tabular}
\label{Simulation parameters}
\end{table}

\begin{figure}[h]
\centering
     \includegraphics[width=.4\textwidth]{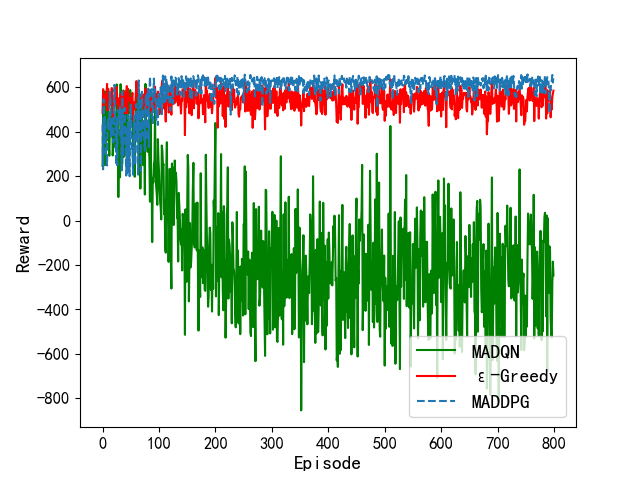} 
\caption{Reward comparison with different algorithms.} 
\label{Reward_Compare}
\end{figure}

\subsection{MADDPG-based Slice Orchestration}
We begin by evaluating the efficiency of the MADDPG-based slice orchestration algorithm. Fig.~\ref{Reward_Compare} presents the convergence comparison for three algorithms: MADDPG, $\varepsilon$-Greedy, and MADQN (Multi-Agent Deep Q-Network). The $\varepsilon$-Greedy algorithm operates by exploring all possible actions in the action space for the given state and then selecting the action with the highest reward after completing the traversal exploration.

The MADDPG curve (blue) converges to approximately $600$ rewards after $160$ episodes. It shows the highest reward and exhibits minimal fluctuations over time. In contrast, the $\varepsilon$-Greedy curve (red) converges to around $500$ rewards, displaying more fluctuations but still demonstrating a generally increasing trend. The MADQN curve (green), however, exhibits significant oscillations and is the only curve that converges to a negative value of $-300$.
Based on the graph, MADDPG appears to learn faster than the other two algorithms, as its curve rises more steeply in the initial stages. While $\varepsilon$-Greedy learns at a slower pace, it still achieves convergence. Compared to $\varepsilon$-Greedy, MADDPG achieves about a $25\%$ higher convergence reward.

This difference can be attributed to the fact that DDPG is specifically designed for continuous action spaces, while DQN, used in MADQN, employs a Q-network to evaluate the value of discrete actions for determining the optimal action. DQN is better suited for discrete action selection. In this scenario, the continuous nature of part of our action space reduces the effectiveness of DQN. Therefore, MADDPG stands out for its stability and faster learning speed in resource allocation for multi-slice orchestration.

\begin{figure}[h]
\centering
     \includegraphics[width=.38\textwidth]{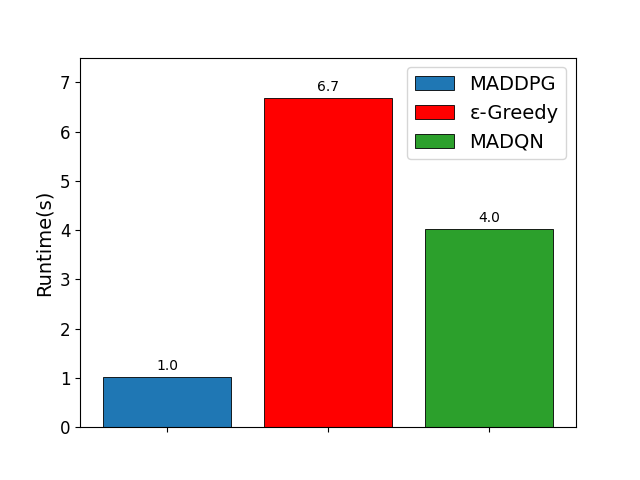} 
\caption{Average time consumption for different algorithms.} 
\label{Runtime}
\end{figure}

Fig.~\ref{Runtime} compares the runtime of MADDPG, $\varepsilon$-Greedy, and MADQN. MADDPG shows the lowest runtime consumption among the three algorithms, indicating its higher computational efficiency. In contrast, the $\varepsilon$-Greedy algorithm requires significantly more time to complete the resource allocation task. This is because $\varepsilon$-Greedy explores all actions in the given action space, which incurs substantial time costs, especially in the complex AAM simulation environment. While $\varepsilon$-Greedy performs reasonably well in terms of reward, its practicality is inferior to that of MADDPG. The considerable difference in runtime could have implications for scalability and real-time applications. Algorithms with lower runtime requirements, such as MADDPG, may be preferable in latency-critical applications.

\begin{figure}
\centering
\subfloat[Total operation cost]{\label{totalOcost}{\includegraphics[width=0.5\linewidth]{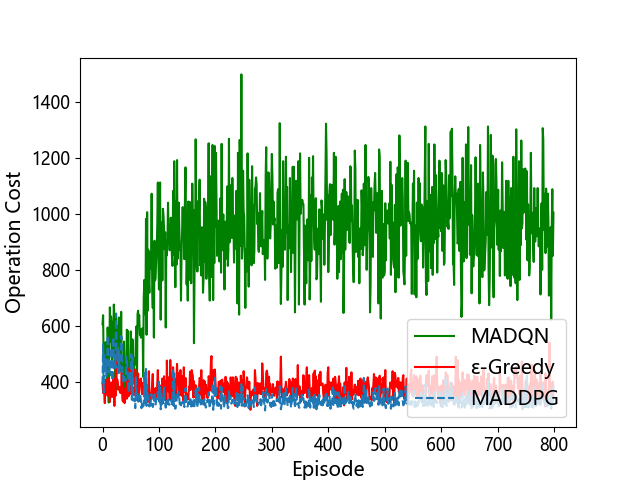}}}
\subfloat[Operation cost from Slice1]{\label{Slice1Ocost}{\includegraphics[width=0.5\linewidth]{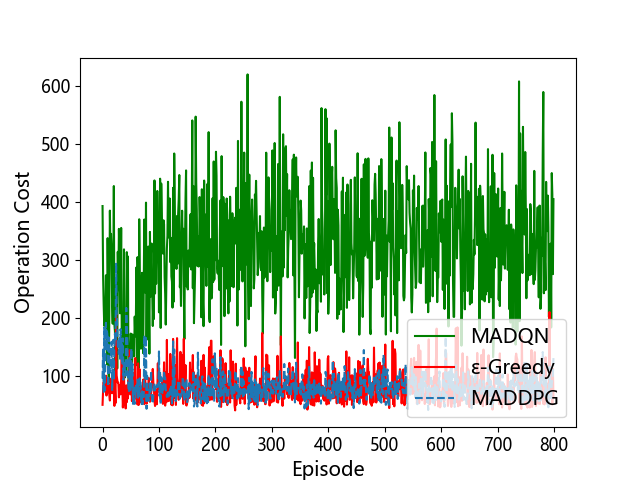}}}
\hfill 
\subfloat[Operation cost from Slice2]{\label{Slice2Ocost}{\includegraphics[width=0.5\linewidth]{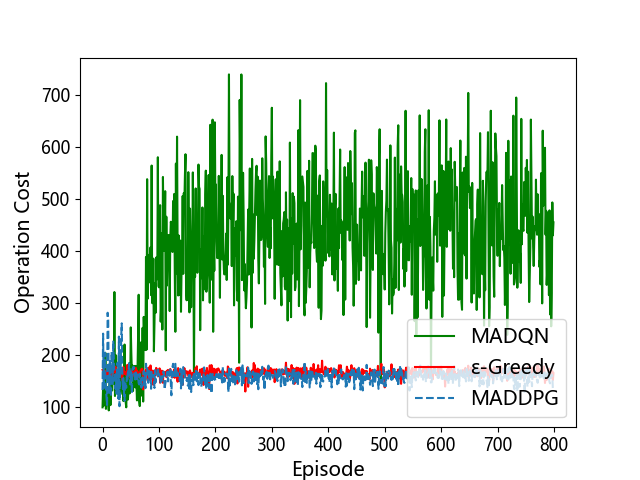}}} 
\subfloat[Operation cost from Slice3]{\label{Slice3Ocost}{\includegraphics[width=0.5\linewidth]{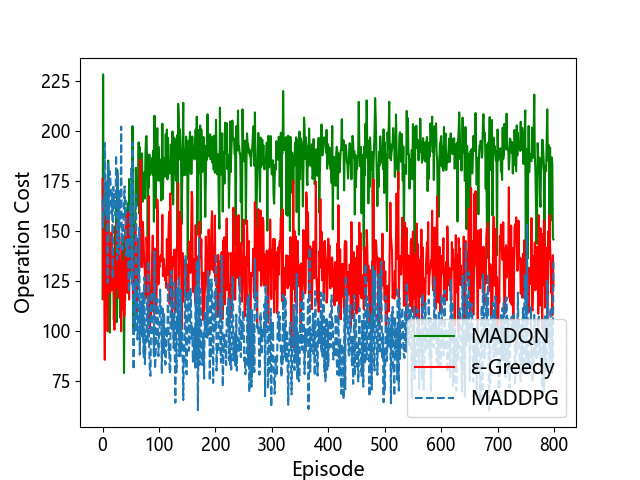}}}
\hfill 
\caption{Operation cost with different algorithms}
\label{dfshghshbvsdjfbo}
\end{figure}

Fig.~\ref{dfshghshbvsdjfbo} illustrates the operation costs associated with different algorithms across three slices (Slice1, Slice2, and Slice3). Fig.~\ref{totalOcost} shows the total operation costs for all slices, while Fig.~\ref{Slice1Ocost}, Fig.~\ref{Slice2Ocost}, and Fig.~\ref{Slice3Ocost} present the individual operation costs for each slice.
In Fig.~\ref{totalOcost}, MADDPG consistently demonstrates the lowest operation costs throughout the observed period. The $\varepsilon$-Greedy algorithm exhibits stable performance, with costs oscillating around an average value. In Fig.~\ref{Slice1Ocost}, MADDPG again leads in cost efficiency, while both $\varepsilon$-Greedy and MADQN show more significant variations. The trends in Fig.~\ref{Slice2Ocost} and Fig.~\ref{Slice3Ocost} mirror that of the total operation cost, with MADDPG providing the most economical solution. Overall, MADDPG achieves the best performance in minimizing operation costs and maintaining stability across all slices.

\begin{figure}
\centering
\subfloat[Total Violation cost]{\label{totalVcost}{\includegraphics[width=0.5\linewidth]{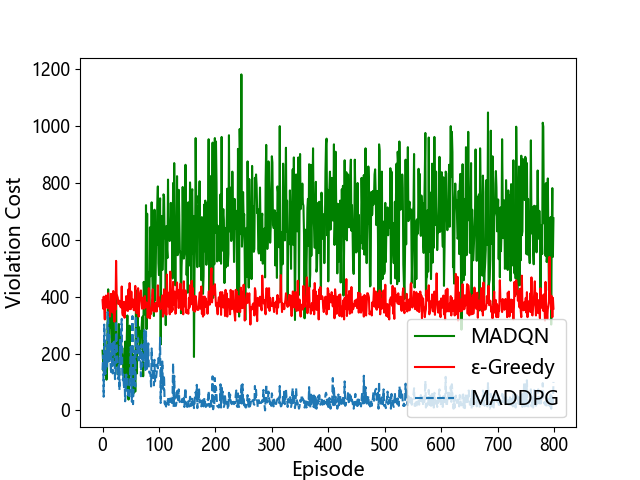}}}
\subfloat[Violation cost from Slice1]{\label{Slice1Vcost}{\includegraphics[width=0.5\linewidth]{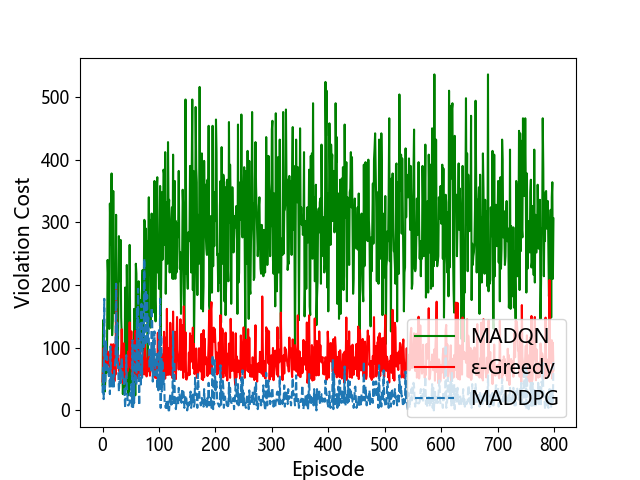}}}
\hfill 
\subfloat[Violation cost from Slice2]{\label{Slice2Vcost}{\includegraphics[width=0.5\linewidth]{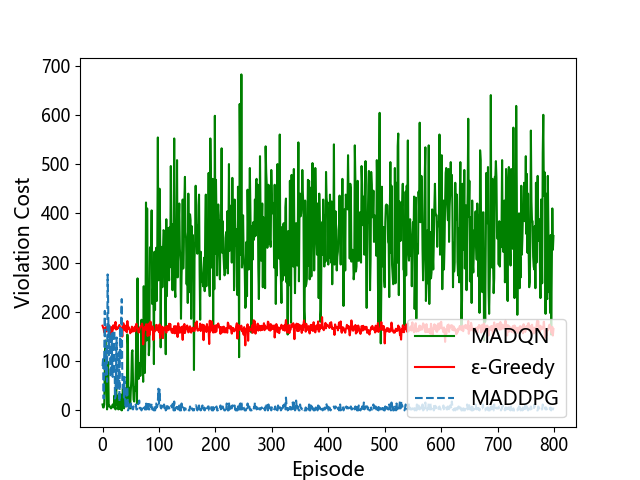}}} 
\subfloat[Violation cost from Slice3]{\label{Slice3Vcost}{\includegraphics[width=0.5\linewidth]{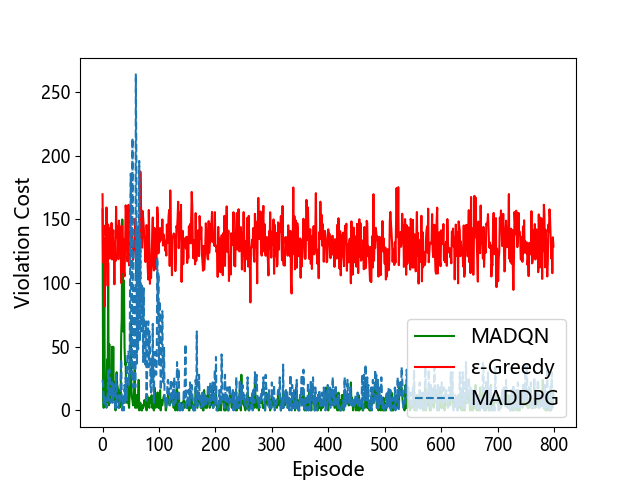}}}
\hfill 
\caption{Violation cost with different algorithms}
\label{dfshghshbvsdjfbs}
\end{figure}

Fig.~\ref{dfshghshbvsdjfbs} illustrates the variation in violation cost (defined as Eq.~(\ref{dfugsueig})) for different algorithms. Specifically, Fig.~\ref{totalVcost} shows the total violation cost for all three slices, while Figs.~\ref{Slice1Vcost}, \ref{Slice2Vcost}, and \ref{Slice3Vcost} present the violation cost for Slice 1, Slice 2, and Slice 3, respectively.

From Fig.~\ref{totalVcost}, MADQN exhibits the highest violation cost, followed by $\varepsilon$-Greedy, while MADDPG shows the lowest violation cost. This trend is also observed in Fig.~\ref{Slice1Vcost} and Fig.~\ref{Slice2Vcost}, indicating that MADQN's resource allocation strategy struggles to meet task requirements. The $\varepsilon$-Greedy algorithm, which selects the action with the highest expected reward for the current state, maintains relatively stable violation costs. In contrast, MADDPG learns from its exploration phase to progressively optimize resource allocation, significantly reducing violation costs and maintaining the lowest violation levels.

Notably, in Fig.~\ref{Slice3Vcost}, MADQN exhibits relatively low violation costs for Slice 3. This behavior may result from MADQN’s unbalanced resource allocation strategy, which likely allocated excessive resources to Slice 3. While this approach reduced violation costs for Slice 3, it led to a significant increase in violation costs for the other two slices, thus decreasing overall resource utilization efficiency.


\begin{figure}
\centering
\subfloat[Total User Satisfaction]{\label{totalSat}{\includegraphics[width=0.5\linewidth]{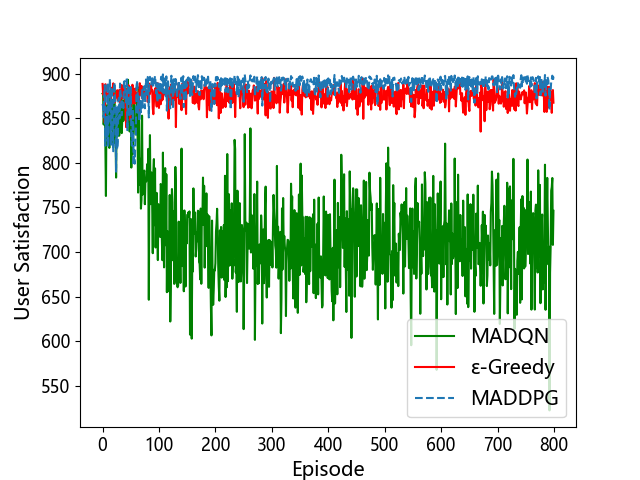}}}
\subfloat[User Satisfaction from Slice1]{\label{Slice1Sat}{\includegraphics[width=0.5\linewidth]{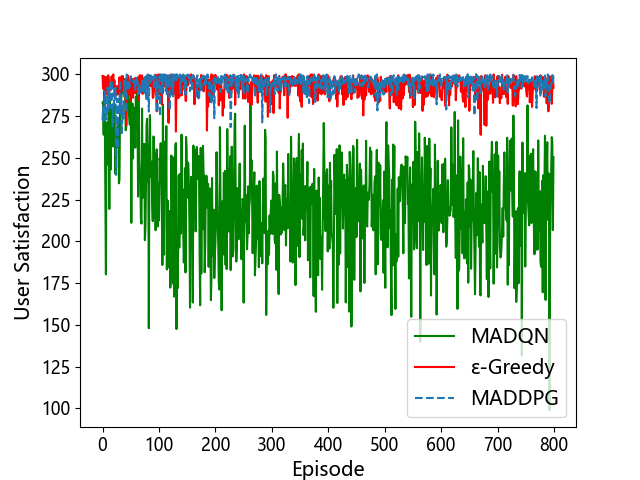}}}
\hfill 
\subfloat[User Satisfaction from Slice2]{\label{Slice2Sat}{\includegraphics[width=0.5\linewidth]{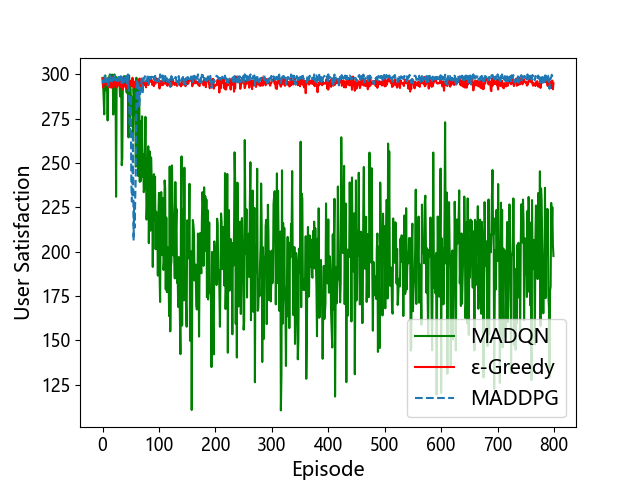}}} 
\subfloat[User Satisfaction from Slice3]{\label{Slice3Sat}{\includegraphics[width=0.5\linewidth]{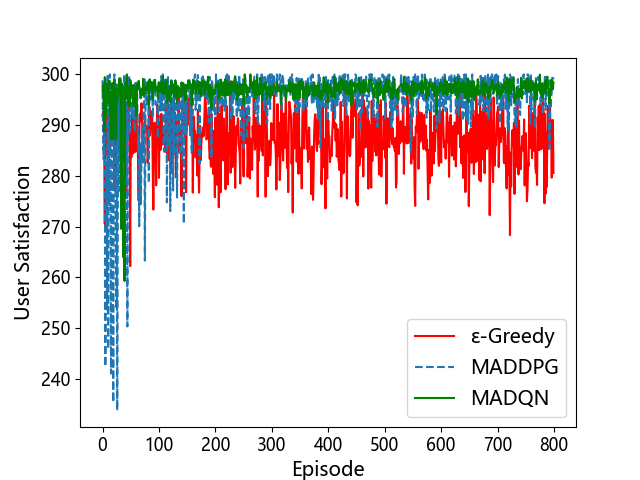}}}
\hfill 
\caption{User Satisfaction with different algorithms}
\label{sdsdfeweweqd}
\end{figure}

Fig.~\ref{sdsdfeweweqd} shows the trends in user satisfaction (defined in Eq.~(\ref{gwetruy})) for different algorithms. Specifically, Fig.~\ref{totalSat} presents the overall system satisfaction curve, while Fig.~\ref{Slice1Sat}, Fig.~\ref{Slice2Sat}, and Fig.~\ref{Slice3Sat} illustrate the individual user satisfaction trends for Slice 1, Slice 2, and Slice 3, respectively.
As seen in Fig.~\ref{totalSat}, \ref{Slice1Sat}, and \ref{Slice2Sat}, MADQN demonstrates significantly lower user satisfaction than both MADDPG and the $\varepsilon$-Greedy algorithm. Additionally, the user satisfaction of $\varepsilon$-Greedy is slightly lower than that of MADDPG. These results show that MADDPG not only achieves higher user satisfaction but also maintains this level of satisfaction over a longer period.

\begin{figure}[h]
\centering
     \includegraphics[width=.4\textwidth]{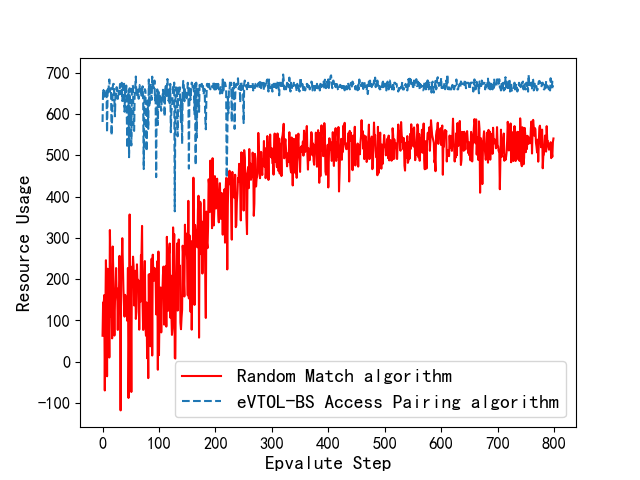} 
\caption{eVTOL-BS matching performance comparison} 
\label{pairing}
\end{figure}

\subsection{eVTOL-BS Pairing}
Fig.~\ref{pairing} compares the performance of the proposed eVTOL-BS pairing algorithm with that of random matching, using the same MADDPG resource allocation strategy. As shown in the figure, the reward obtained through the proposed access pairing algorithm is significantly higher than that of random eVTOL-BS-Slice matching. This highlights the effectiveness of the proposed algorithm in eVTOL-BS-Slice matching.

\begin{figure}
\centering
\subfloat[Total resource consumption]{\label{resource}{\includegraphics[width=0.5\linewidth]{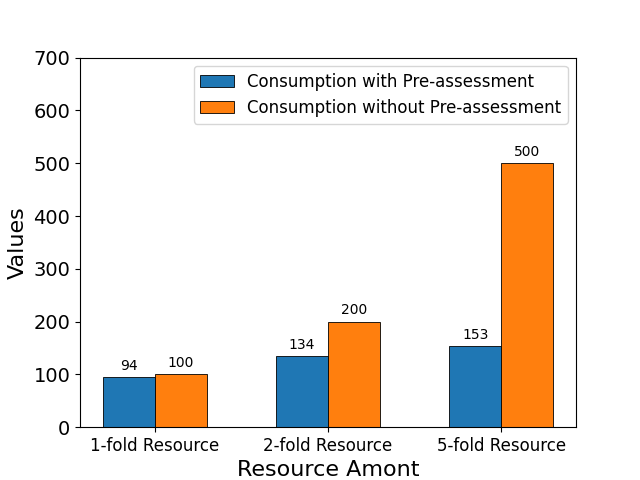}}}
\subfloat[Resource consumption per episode]{\label{evaluate}{\includegraphics[width=0.5\linewidth]{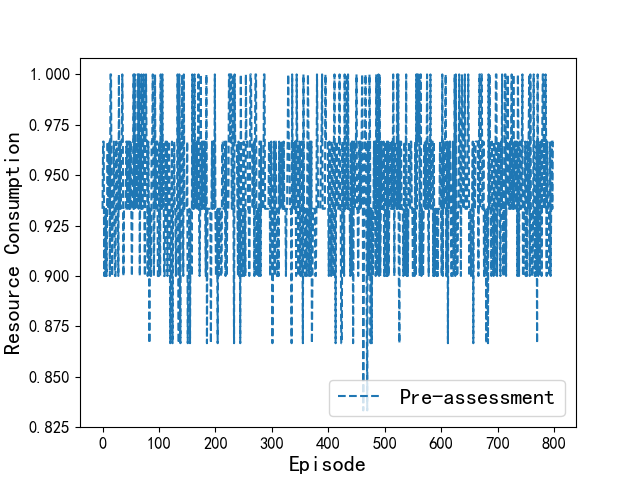}}}
\hfill 
\subfloat[Operation cost comparison]{\label{op_assessment}{\includegraphics[width=0.5\linewidth]{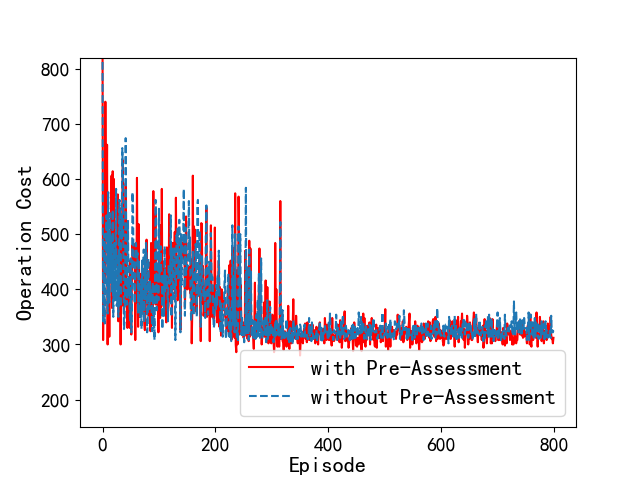}}}
\subfloat[Violation cost comparison]{\label{vio_assessment}{\includegraphics[width=0.5\linewidth]{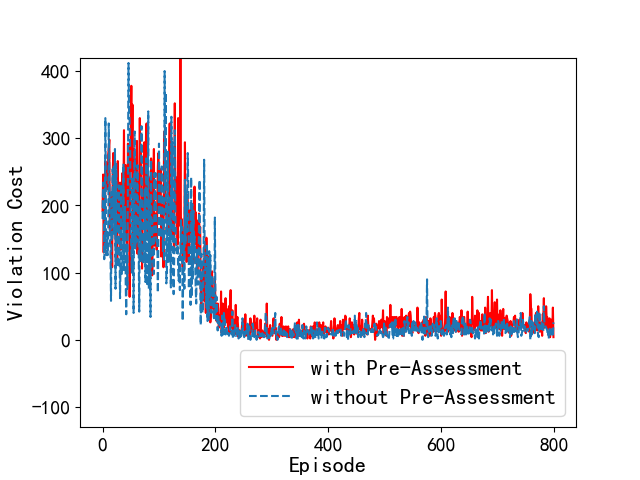}}}
\hfill 
\caption{Performance gain from the pre-assessment algorithm}
\label{ajdshfjarfuyuwqo}
\end{figure}

\subsection{Resource Pre-assessment}

Fig.\ref{ajdshfjarfuyuwqo} illustrates the performance gains achieved with the pre-assessment algorithm. In this simulation, the total available resources are normalized to $1.0$. As shown in Fig.\ref{resource}, incorporating the pre-assessment algorithm results in a $6\%$ reduction in resource consumption with 1-fold total resources. When the available resources are doubled (2-fold), resource consumption decreases by approximately $33\%$ compared to the baseline without the pre-assessment algorithm. The difference becomes even more pronounced with 5-fold resources, where the pre-assessment algorithm reduces consumption by $69\%$.

These results demonstrate that the pre-assessment algorithm effectively improves resource usage with adequate resource allocation. Furthermore, Fig.~\ref{evaluate} provides a detailed view of resource consumption for each episode under the pre-assessment algorithm, showing that it avoids exhausting all available resources in most cases.

Fig.~\ref{vio_assessment} and Fig.~\ref{op_assessment} indicate that operation and violation costs remain largely unchanged with or without the pre-assessment algorithm. This suggests that the pre-assessment algorithm does not increase operational or violation costs, while significantly reducing resource consumption. These findings validate the effectiveness of the proposed algorithm.

\begin{figure}[h]
\centering
     \includegraphics[width=.4\textwidth]{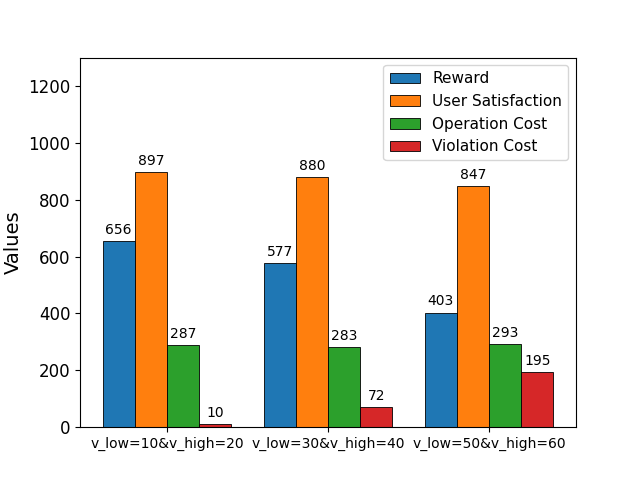} 
\caption{Performance with different eVTOL velocities at high/low layers.} 
\label{speed_comparison}
\end{figure}

\subsection{Impact of eVTOL Mobility}
To assess the impact of eVTOL mobility on simulation performance, Fig.~\ref{speed_comparison} presents the variations in Reward, Satisfaction, and Operation/Violation Cost for MADDPG under three different velocity conditions: $v_{low} = 10 \ \& \ v_{high} = 20$, $v_{low} = 30 \ \& \ v_{high} = 40$, and $v_{low} = 50 \ \& \ v_{high} = 60$.
Here, $v_{low}$ represents the eVTOL velocity in the low-speed layer, while $v_{high}$ denotes the velocity in the high-speed layer. The velocity difference between the high-speed and low-speed layers is fixed at $20$. As eVTOL velocity increases, both Reward and Satisfaction decrease, while Operation/Violation Costs rise. This is because higher velocities degrade the channel conditions between eVTOLs and ground BSs and negatively impact beam tracking performance, ultimately reducing user satisfaction and increasing costs.

\begin{figure}[h]
\centering
     \includegraphics[width=.4\textwidth]{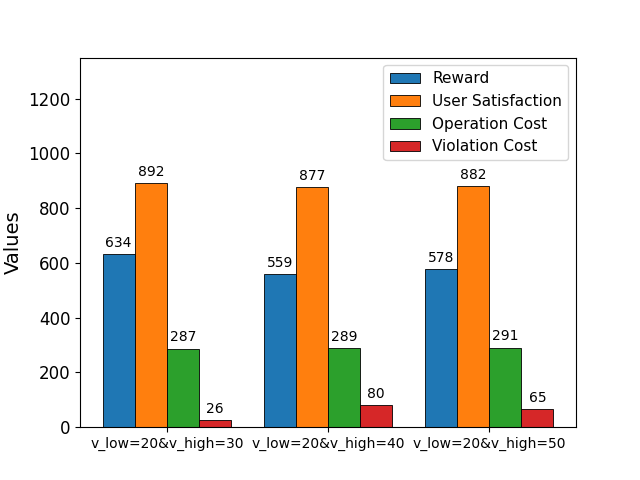} 
\caption{System performance v.s. velocity difference between the low-speed and high-speed layers.} 
\label{speed_change_high_layer}
\end{figure}

To further investigate the effect of the velocity difference between the low-speed and high-speed layers on system performance, Fig.~\ref{speed_change_high_layer} shows a bar chart comparing performance when the low-speed layer's velocity is fixed at $20$ and the high-speed layer's velocity is varied.
As the velocity of the high-speed layer increases, both Reward and Satisfaction decrease, while Operation/Violation Cost increases. This occurs because higher eVTOL speeds in the high-speed layer cause eVTOLs to leave the coverage area of the current BS more quickly, resulting in poorer channel conditions and beam tracking issues. These effects ultimately lead to a decline in overall system performance.


\section{Conclusion}
This paper proposed a low-altitude intelligent network slicing framework for AAM systems, specifically designed to address the unique task offloading challenges in eVTOL networks. By leveraging intelligent network slicing, the proposed algorithm dynamically allocates heterogeneous resources—such as bandwidth, beam alignment, and computing—based on real-time flight patterns and task requirements of eVTOLs.
In addition, we novelly introduce a slice admission control module that pre-schedules eVTOL-BS-Slice pairings and allocates available resources, thereby enhancing resource utilization and reducing consumption. Building on the outcomes of this module, the proposed MADDPG algorithm further optimizes task offloading within a layered AAM system. It achieves superior performance in minimizing operation and violation costs, while also improving offloading efficiency.
This work paves the way for low-altitude intelligent network design, with fully considering eVTOL mobility and heterogeneous resource allocations in a layered airspace.




\appendices
\footnotesize
\bibliography{biblio}

\end{document}